# An Experimental Study of Satisfaction Response: Evaluation of Online Collaborative Learning

**Xusen Cheng[1], Xueyin Wang[1], Jianqing Huang[1], Alex Zarifis[2]**
[1]University of International Business and Economics, China, [2]University of Mannheim, Germany

## Abstract

On the one hand, a growing amount of research discusses support for improving online collaborative learning quality, and many indicators are focused to assess its success. On the other hand, thinkLets for designing reputable and valuable collaborative processes have been developed for more than ten years. However, few studies try to apply thinkLets to online collaborative learning. This paper introduces thinkLets to online collaborative learning and experimentally tests its effectiveness with participants' responses on their satisfaction. Yield Shift Theory (YST), a causal theory explaining inner satisfaction, is adopted. In the experiment, 113 students from Universities in Beijing, China are chosen as a sample. They were divided into two groups, collaborating online in a simulated class. Then, YST in student groups under online collaborative learning is validated, a comparison study of online collaborative learning with and without thinkLets is implemented, and the satisfaction response of participants are analyzed. As a result of this comparison, YST is proved applicable in this context, and satisfaction is higher in online collaborative learning with thinkLets.

An Experimental Study of Satisfaction Response: Evaluation of Online Collaborative Learning
Xusen Cheng, Xueyin Wang, Jianqing Huang, Alex Zarifis

# Introduction

Collaborative learning is a method for students working together in a small group toward the same goal (Prince, 2004). Many researchers have discovered that collaborative learning positively impacts the learning process since group diversity evokes a re-thinking of the intention of every participant (Vygotsky, 1978). With the rapid development of science and technology, the transition from the traditional class to an online classroom and online learning is getting much easier. This also applies to online collaborative learning. Hence, online collaborative learning, the computer-supported version of in-class collaboration (Ku et al., 2013), attracts great attention. According to Swan et al. (2006), online collaborative learning holds remarkable potential to support learning. It not only provides all participants with the chance to have an equal voice, but also reflects students' contributions and writing (Gunawardena and Zittle, 1997). Therefore, it is believed by researchers that online collaborative learning is a good way to improve the quality of online course designs (Biasutti, 2011).

There are many technologies and features we use, such as web-based systems, for online collaborative learning (McGreal and Elliott, 2008). However, efficiency of it can't be taken for granted. Students may to some degree suffer from frustration in online collaborative learning because negative emotions and stress are evoked by past experiences or by collaborating with people they do not know well (Capdeferro and Romero, 2012). As a result, if new tools and supports are introduced in online collaborative learning, it is truly vital to assess whether it is successful under representative indicators. To determine successfulness of online collaborative learning, trust among team members, team peformance, and so on are assessed(Cheng & Macaulay, 2014; Cheng et al., 2013a; Lee et al., 2011). Satisfaction, as one of the most important indicators, is widely discussed due to its significant relation with online events continuance. It is measured in many fields, such as e-service quality (Al-Nuaimi et al., 2013), virtual organizations (Taylor et al., 2013), information system success (DeLone and McLean, 1992; Briggs et al., 2008; Kang and Lee, 2010) and so on, especially in online collaborative learning (Sun et al., 2008; Lee et al., 2011; Zhu, 2012; Ku et al., 2013; Chua and Montalbo, 2014).

For most collaborative groups, it can be a thorny problem to design effective processes (de Vreede et al., 2006), especially for student groups without process designing experience. Researchers began to find ways to support groups to yield better process. Group Support System (GSS) with *thinkLet* concepts, which is derived from collaboration engineering, is developed for collaborative process design (Briggs et al., 2003). GSS offers support for optimization of co-production processes (Briggs et al., 2010), and thinkLet is a packet of collaboration patterns to be applied by practitioners to create repeatable collaboration processes (Kolfschoten et al., 2006). They together make it easier to implement collaboration. However, whether thinkLets are introduced to online collaborative learning and whether the particular selection and integration of them embedded in GSS could create more value still needs to be verified by indicators of measurement, such as satisfaction.

However, satisfaction findings are varied, complex, and even contradictory. In online collaborative learning, interaction is sometimes regarded as the indicator of learners' satisfaction (Kuo et al., 2014); Human factors are also found important in learners' satisfaction (Alshare et al., 2011). These complicated findings are relied on in the measurement of satisfaction responses in online collaborative learning. However, when thinkLets are introduced to online collaborative learning, more suitable and scientific causal theory should be applied. Yield Shift Theory (YST) is a new causal theory to offer more complete explanations for satisfactory phenomenon in the information system domain (Briggs et al., 2008). It implies an inner mechanism of satisfaction responses. Therefore, the selection of YST to test satisfaction responses makes the study more precise and representative. Recently, investigators have examined the validation of the theory (Sindhav et al., 2011; Briggs et al., 2014). Despite this, YST still needs to be verified under another scenario: student groups to be reasonably applied to this study.

So far, there has been very little discussion of students' satisfaction responses of collaboration engineering applied to the field of online collaborative learning, adopting the causal theory YST to do scientific evaluation. In this study, student groups collaborate online with the help of thinkLets and GSS, which is also the context and discipline of the study. Whether GSS with thinkLets is a success for online collaborative learning or whether it could enhance group members' satisfaction still needs to be confirmed. This paper attempts to address the knowledge shortage with the following questions:

1. Is YST sound when applied in student groups when implementing online collaborative learning?

2. In student groups, do groups that collaborate online with thinkLets feel more satisfied than groups without thinkLets under the support of GSS?

In order to answer these questions, this paper would first verify YST in student groups. If YST is sound in this context, then question 2, to verify the success of introducing thinkLets and GSS to online collaborative learning, can be answered by the adoption of YST to test satisfaction responses of student groups with or without thinkLets in online collaborative learning.

**Online Collaborative Learning**

Computer-mediated collaboration in learning has been given extensive attention from researchers over decades (Curtis and Lawson, 2001). For instance, trust development of participants in online collaboration, tools or supports for it, and the relationship of its factors are widely discussed (Cheng and Macaulay, 2014; Lee et al., 2011). As for tools and supports, a variety of technologies are used to offer support for online collaborative learning. For example, firstly, in 1997, a web-based collaborative learning tool, Virtual-U was presented (Harasim et al., 1997). As a second example, from 2005, a shared document-based annotation tool to support collaborative learning was presented and its usefulness was empirically tested (Nokelainen et al., 2005). Recently, an Emoticon support tool in online collaborative learning was developed to facilitate peers' feedback (Lim et al., 2012). However, whether these tools and supports for online collaborative learning are useful and scientific is hardly determined in different aspects.

In the field of online collaborative learning, the focus is still extensively on satisfaction. Sun et al. (2008) concluded that a satisfaction framework includs six dimensions in online learning: learner, instructor, course, technology, design, and environment. Then, the relationship between satisfaction, outcome, and student perception of support was examined (Lee et al., 2011). Later on, students' satisfaction of online collaborative learning was analyzed from a cross-cultural perspective, showing differences of satisfaction between students from different countries (Zhu, 2012). Interaction was then regarded as a remarkable indicator of satisfaction (Kuo et al., 2014). However, most of these studies utilize different theories to evaluate satisfaction. It is obvious that different domains and contexts require suitable frameworks respectively to test satisfaction responses. Hence, in the field of online collaborative learning, reliable and suitable causal theory should be adopted in research.

In this study, thinkLets are introduced to online collaborative learning in design collaboration process, and a self-developed system *Discussion* with four typical *thinkLets* embedded, serving as GSS, is used. Then, satisfaction responses of participants are analyzed.

**ThinkLets**

Utilizing a Group Support System (GSS) to support online collaboration is a significant breakthrough in the field of thinking pattern creation to make collaboration processes more intelligent. Current GSS provides tools for collaborative ideation, reducing, clarifying, organizing, and evaluating (Briggs et al., 2010). These collaborative patterns depend on the selecting of thinkLets.

A thinkLet is the smallest unit which is documented and provided to create a known pattern of collaboration, which makes it transferable, reusable, and predictable to design a collaboration process (Kolfschoten et al., 2006; de Vreede et al., 2009). ThinkLets offer elaborative guidance to facilitators, helping them facilitate their groups better. ThinkLets were proposed by researchers in 2001, and a thinkLet has three components: tool, configuration, and script. Tool means specific software and hardware technology which provides support for creating a pattern of thinking; configuration is the specifics which explain how the software and hardware were configured to create a pattern of thinking; and script is the description and instruction of teaming strategies, facilitation approaches, and activity processes (Briggs et al., 2001). It states everything about what facilitator will say to others, what others do, and how to do it over the whole process.

Among more than 60 thinkLets to be chosen, there are several thinkLets commonly used. Instances are DirectedBrainstorm whose purpose is to generate broad and highly creative ideas by group members, PopcornSort to classify created ideas into categories, BucketWalk to evaluate results of each category to determine whether they are appropriate, and StrawPoll to evaluate concepts with several criteria (Noor et al., 2008; de Vreede et al., 2009). These four thinkLets are usually adopted by researchers (Azadegan et al., 2013; Cheng et al., 2013b).

**Yield Shift Theory**

The Yield Shift Theory (YST) is a cognitive theory, adding an affective component to explicate the inner response and nature of satisfaction (Briggs et al., 2008). When group members use GSS with thinkLets to implement online collaboration, YST can be adopted to test their satisfaction response to evaluate whether GSS with thinkLets is a success in online collaboration.

In each collaboration, there is an active goal, which is a subset of goals that can be assessed by human beings at one moment. People's *utility*, a sense of goodness, worth, or value, is assessed, underlying the assumption that the level of utility to a given active goal is automatically ascribed to a cognitive mechanism. Similarly, perceived *yield* and the *likelihood* that an active goal may be attained are subconsciously assessed. YST advocates that perceived yield is a multiplicative function of utility, where utility is a causal construct and yield is a consequent construct, with likelihood moderating (Briggs et al., 2008).

However, it is not yield shift at a given moment but yield itself that leads to a response of satisfaction. Under the circumstances, utility and likelihood, as well as the active goal set, could change. That yield shift can be automatically detected and that affective responses derive from the shift are assumptions based on YST as well. The whole process or mechanism of YST proposed by Briggs et al. (2008) is presented in Figure 1.

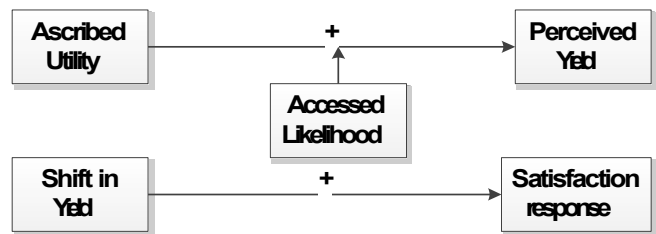

*Figure 1*. Mechanism of YST: Perceived Yield is a function of Ascribed Utility to an active goal set with moderator Assessed Likelihood; Satisfaction is a function of Shift in Yield.

YST is an improved mechanism to explain satisfaction response. Its scientificity was tested by Sindhav et al. (2010) with *nostalgia effects*. Ten effects such as anticipation effects, confirmation effects, disconfirmation effects, attenuation effects, etc. can be explained by the new theory (Shiv et al., 2000; Bhattacherjee, 2001; Suh et al., 1994; McKinney et al., 2002). The empirical field study of YST was conducted by Briggs et al. (2014).

However, so far, satisfaction has not been tested by this scientific theory in online collaborative student groups with thinkLets. An experimental comparison study in which student groups are reasonably applied to this study should be conducted to better understand satisfaction responses of online collaborative learning with or without thinkLets.

# Study Methodology

In this paper, student groups were tested in the experiment. Learners' satisfaction responses of thinkLets and GSS processes in online collaborative learning were measured by YST to indicate whether the attempt of introducing thinkLets and GSS was a success.

**Dependent Variables**

Two dependent variables are used in the study: *Satisfaction-with-Process* (SP), a satisfaction response related to collaboration method and procedure, and *Satisfaction-with-Outcome* (SO), a satisfaction response related to result of meetings. These two dependent variables are measured with a five-item, five-score Lickert scale which has been verified in previous studies (Reinig et al., 2009; Briggs et al., 2008; Briggs et al., 2014) respectively. Items of SP are:

*1.     I feel satisfied with the way in which today's meeting (discussion) was conducted.*

*2.     I feel good about today's meeting (discussion) process.*

*3.     I liked the way the meeting (discussion) progressed today.*

*4.     I feel satisfied with the procedures used in today's meeting (discussion).*

*5.     I feel satisfied about the way we carried out the activities in today's meeting (discussion).*

Items of SO are:

*1.  I liked the outcome of today's meeting (discussion).*

*2.  I feel satisfied with the things we achieved in today's meeting (discussion).*

*3.  When the meeting (discussion) was over, I felt satisfied with the results.*

*4.  Our accomplishments today give me a feeling of satisfaction.*

*5.  I am happy with the results of today's meeting (discussion).*

**Interdependent Variables**

Two variables in YST, Likelihood Shift (LS), which indicates shift in Assessed Likelihood, and Utility Shift (US), which indicates shift in Ascribed Utility, are still used in the study. Both of them are measured with a four-item, five-score semantic anchor scale which has been verified in previous studies (Sindhav, 2011; Briggs et al., 2014). Items of LS are:

*1.  The meeting made it (more/less) likely that I would attain something I want.*

*2.  Because of the meeting, I am (more/less) likely to succeed on something I care about.*

*3.  I am (more/less) likely to attain my goals because of this meeting.*

*4. Due to this meeting I am (more/less) likely to get what I want.*

Items of US are:

*1. I got (more/less) from the meeting than I had anticipated.*

*2. I benefited (more/less) from this meeting than I expected.*

*3. The meeting did (more/less) good for me than I thought it would.*

*4. I gained (more/less) from the meeting than I believed I would.*

**Participants**

113 participants—54 males and 59 females—took part in the study, all of whom were undergraduate students in universities of China in Beijing, including University of International Business and Economics, Beijing University of Posts and Telecommunications, Beijing Jiaotong University, and China Agricultural University. They indicated that they have never used thinkLets before. Each of them provided personal information relevant to the study.

At first, 113 students were divided into two groups randomly, group A with 59 members and group B with 54 members. Group A, aging from 18 to 22 (average age$_A$=19.915), consisted of 28 males and 31 females. Group B, aging from 18 to 21 (average age$_B$=19.981), consisted of 26 males and 28 females. Group A and Group B were divided into several smaller groups. Each small group was assigned a facilitator. We demonstrated how to operate the online collaboration system and explained the steps of a collaboration process with scripts. The facilitator was the one who mastered the meeting process and timeframe. Other group members collaborated under the instructions and guidance of the facilitator. In group A, Brainstorming, PopcornSort, BucketWalk, and StrawPoll were adopted with thinkLet scripts guiding and with a self-developed meeting system *Discussion* assisting to discuss a specific topic within 20 minutes. In this experiment, we simulated an online class and set *what kind of system to develop for the final task of the course: Database System* as the discussion topic. Group B discussed the same topic within the same time with QQ Group assisting. Discussion is a system with these exact four patterns of collaboration, while QQ Group is what students are accustomed to use when they freely discuss in groups (Li, 2012). Apart from the use or not of thinkLets, there is no difference in using the two systems Discussion and QQ Group. The procedure of the study is presented in Figure 2.

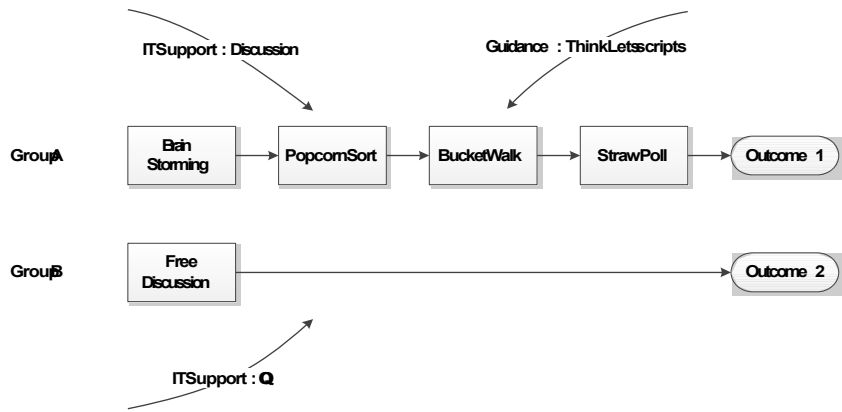

*Figure 2.* Procedure of collaboration in Group A and Group B.

**Data Collection**

After the experiment, all 113 participants were asked to take two minutes to respond to the one-page instruments. Participants were told that the instruments were anonymous. After finishing the instruments, participants were asked to be interviewed voluntarily. All instruments were valid in spite of several items missing values. So, there were 113 instruments, and 10 participants in Group A and 8 participants in Group B who accepted the interview.

# Results

**Questionnaire Data Analysis**

Since participants were divided into two groups to make the comparison, there should be experimental control over Group A and Group B. The similar participants' composition and the experimental control of the two groups assured that there was no significant difference of participants in terms of demographic characteristics or experimental condition that could affect the outcome of the study in the two groups.

**Significant Correlations**

The data shows significant statistically positive correlation between US and SP, US and SO, LS and SP, LS and SO, which is presented in Table 1. For US and SP, correlation is .793, p<0.001. For US and SO, correlation is .738, p<0.001. For LS and SP, Ccrrelation is .740, P<0.001. For LS and SO, correlation is .709, p<0.001. This suggests that the more shift participants reported, the more satisfaction they responded.

| | Correlations | |
|---|---|---|
| | Utility Shift | Likelihood Shift |
| Satisfaction-with-Process | .793** | .740** |
| Satisfaction-with-Outcome | .738** | .709** |
| N=113 | P<0.001 | |

Table 1 Significant Correlations between US and Satisfaction Response, LS and Satisfaction Response.

**Higher Satisfaction with ThinkLets**

The mean value of responses to each item in the scale indicates how people generally respond to each factor; the standard deviation is used to address the degree that responses to items of each scale vary. However, in some cases, there are several items missing values. These blanks are replaced by the mean value computed with completed items. Table 2 shows the result of the questionnaire of the two groups.

In essence, the higher the mean of the factor is, the greater people rate it, and the lower the standard deviation is, the greater the consensus of people's responses is. Hence, some of the most influential factors can be identified by analysis combining mean values and standard deviation. The statistical data presented in Table 1 indicates that there are differences of satisfaction response between Group A and Group B.

| | Mean | | | | Standard deviation | | | |
|---|---|---|---|---|---|---|---|---|
| Groups | SP | SO | US | LS | SP | SO | US | LS |
| A | 3.996 | 3.925 | 2.945 | 3.110 | 0.831 | 0.838 | 1.508 | 1.581 |
| B | 3.525 | 3.629 | 1.921 | 2.083 | 0.963 | 0.935 | 2.080 | 2.129 |

Table 2 More Satisfaction in Group A ($N_A$=59), Less in Group B ($N_B$=54), Indicating from Questionnaire Responses, Scale 1 to 5, (1) Strongly Disagree, (5) Strongly Agree

Mean values of SWP, SWO, US and LS in Group A are higher than that in Group B, while standard deviations are lower. This means most students collaborating with thinkLets embedded in the collaborative software reported more utility shift and likelihood shift, and they felt more satisfied with meeting process and outcome.

**Significant Difference in SP, SO, US, LS**

Moreover, T-test was used to further validate the differences of satisfaction between the two groups. Results are presented in Table 3. Take SP and SO, for example: Sig. (SP) > 0.05 and Sig. (SO) > 0.05. Homogeneity of variances is obviously indicated. Two-tail Sig. (SP) < 0.05 and two-tail Sig. (SO) < 0.05, which means differences of Satisfaction-with-Process and Satisfaction-with-Outcome are significant between Group A and Group B. As for US and LS, Sig. (US) <0.05 and Sig. (LS) <0.05, showing that variances are not homogeneous. Second line of US, LS data should be examined. While two-tail Sig. (US) <0.05 and two-tail Sig. (LS) < 0.05, it can be concluded that Utility Shift and Likelihood Shift are significantly different as well between Group A and Group B.

|    | Levene test | | T-test | | | | | Confidence interval (95%) | |
|----|-------|------|--------|---------|------------------|-----------------|------------------|----------------|----------------|
|    | F     | Sig. | t      | df      | Sig. (two-tail)  | Mean difference | Standard errors  | Upper bound    | Lower bound    |
| SP | 1.967 | .164 | -3.296 | 111     | .001             | -2.35342        | .71403           | -3.76831       | -.93853        |
|    |       |      | -3.266 | 102.260 | .001             | -2.35342        | .72059           | -3.78267       | -.92418        |
| SO | .447  | .505 | -2.127 | 111     | .036             | -1.47897        | .69521           | -2.85658       | -.10136        |
|    |       |      | -2.113 | 105.123 | .037             | -1.47897        | .69983           | -2.86658       | -.09136        |
| US | 4.554 | .035 | -3.278 | 111     | .001             | -4.09448        | 1.24895          | -6.56935       | -1.61960       |
|    |       |      | -3.232 | 95.338  | .002             | -4.09448        | 1.26696          | -6.60960       | -1.57935       |
| LS | 7.172 | .009 | -3.078 | 111     | .003             | -4.10734        | 1.33463          | -6.75199       | -1.46269       |
|    |       |      | -3.034 | 95.403  | .003             | -4.10734        | 1.35382          | -6.79486       | -1.41983       |

Table 3 Significant Difference in SP, SO, US, LS from T-test Result in Group A and Group B

**Interview Data Analysis**

There were 10 participants from Group A and 8 participants from Group B interviewed after the meeting. The interview questions are listed in Table 4. These questions are compatible with SO, SP, US, LS.

1. Is it worthwhile to spend time collaborating?
2. Are you satisfied with the process? Why?
3. Will former bad/good collaboration experiences have effect on this one?
4. Are you satisfied with the outcome? Why?
5. Do you feel thinkLets bring you more?
6. Do you feel thinkLets likely to give you better outcome?

Table 4 Interview Questions

To handle interview data, we gave each interviewee a detailed number in Group A ($A_1$ to $A_{10}$) and Group B ($B_1$ to $B_8$) first and we finally extracted critical statements and keywords in the coding of the interview and drew the conclusion.

Eight of the ten participants in Group A reported more US and LS, and higher SP and SP, such as:

> $A_1$: "I feel satisfied with the process, especially Brainstorming. It makes me think independently when collaborating online, and the process makes the meeting smoother."
>
> $A_2$: "Maybe for me, the result is not exactly what I proposed, but it's the best result for the group. Anyway, I feel good about it."
>
> $A_4$: "I gain more utility since it saves us much time to get a consistent conclusion. It definitely improves effectiveness."
>
> $A_{10}$: "StrawPoll makes it possible for more fairness, since it's statistically scientific. When it was imbedding in a collaborative software, the result is more vivid and clear, and because of that, everyone tends to be more satisfied to the result."

However, in Group B, only one participant reported higher SO. Others felt no significant difference in aspects of US, LS, SP, and SO, such as:

> $B_1$: "I don't believe we can gain more with the similar mode of discussion. It takes much time since most of people are inattentive."
>
> $B_4$: "Actually, I don't feel good about the process. We didn't control time well, and we sometimes are off the topic, since it's online, you know. Emm… Thus you can think about the result."
>
> $B_5$: "There are so many defferent ideas. It's hard to decide. I believe it's same difficult for other group members."
>
> $B_8$: "Without some criterions to evaluate ideas, everyone seems to be passionless to agree with others."

In the statements of the interviews, we firstly marked all key words and phrases related to commenting on the online collaborative learning. If there was no apparent key words in a comment-like sentence, we concluded them into words or phrases respectively. Then, we listed all original and concluded words and phrases, counting the times they were mentioned. After analyzing the interview data, we found out that there were several keywords that are related to satisfaction mentioned frequently. The result, from the more frequently mentioned to less frequently mentioned, is shown in Figure 3.

70

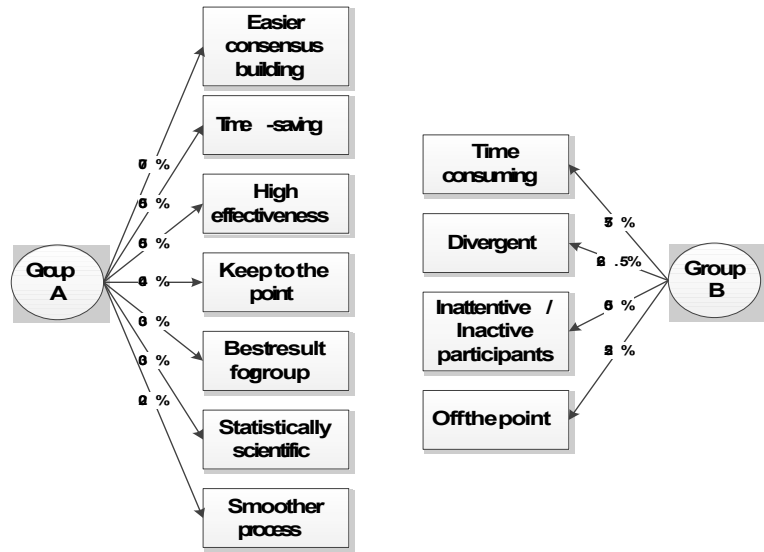

*Figure 3.* Analysis of interview data in Group A and Group B: percentage represents the frequency of each keyword mentioned by participants.

From the Analysis, we can find that:

- With thinkLets in online collaborative learning, yield shift will be more which leads to higher satisfaction.

- ThinkLets help to improve quality of online collaborative learning in aspects of time-saving, consensus building, effectiveness, scientificity, and so on.

# Discussion

**Consistency of YST Result in Online Collaborative Student Groups**

Utility Shift and Likelihood Shift correlate positively with satisfaction responses. Hence, under the conditions of this study, the observed relationships between yield shifts and satisfaction responses were consistent with what was proposed by YST theoretically. The more yield shifts were reported, the more satisfied participants tended to be, and the less yield shifts were reported, the less satisfied participants tended to be. This answered Research Question 1. YST is sound when applied to student groups in online collaborative learning. Only if YST is proved valid in this context, can analysis of Question 2 become meaningful.

**Group Collaborating with ThinkLets Reports Higher Satisfaction**

During online collaborative learning in the group that participants collaborated with thinkLets, higher scores in yield shifts and satisfaction responses were reported. In the other group, where participants collaborated without thinkLets, lower scores were reported. Students felt more satisfied with the process and the final outcome of online collaboration when they used thinkLets. Many participants mention "time saving", "consensus outcome", "effective process", and so on with the help of thinkLets in collaborative system, and they feel good about it. This answered Research Question 2. Groups collaborating with thinkLets feel more satisfied than groups collaborating without thinkLets in online collaborative learning.

**Implication for Practice**

There are several practical implications from these findings. Using thinkLets to design collaborative process contributes to a higher satisfaction response from participants in online collaborative learning. In designing collaborative software, there are more aspects worth attention. Not only the basic function of the software, but also supportive processes leading to a better outcome and user satisfaction need to be taken into consideration.

**Implication for Research**

The indicator satisfaction was proposed a long time ago. It has been the focus of lots of researchers, and it was measured in many fields like online learning, e-service quality, virtual organizations, and information system success (Chua and Montalbo, 2014; Al-Nuaimi et al., 2013; Taylor et al., 2013; Kang and Lee, 2010). However, satisfaction findings are varied, complicated, and even contradictory. The measurement of satisfaction responses often relied on these complicated findings, rather than scientific and causal theory. YST is the newly developed causal theory that tends to explain the inner response and nature of satisfaction effects (Briggs et al., 2008). This study adopted YST, and it produced results which corroborate the findings of a great deal of the previous work in this field (Sindhav et al., 2011; Briggs et al., 2014). In this paper, YST was verified in student groups, indicating that YST is useful in one more occasion. Furthermore, YST was introduced to the field of online collaborative learning. This paper adopted YST to test students' satisfaction when they collaborated online with GSS. The observed satisfaction responses are consistent with YST propositions, so the results suggested that YST may be a helpful perspective to understand satisfaction in online collaborative learning.

On the one hand, there has been much effort devoted in the research of thinkLets since Briggs proposed it (Briggs et al., 2001). A framework for designing collaborative process, models for collaboration system design, theories to assess collaboration success, and so on have been gradually developed (de Vreede et al., 2009; Briggs, 2006; Briggs et al., 2008). On the other hand, online collaborative learning has been discussed a lot, such as tools and supports to improve quality of learning (Harasim et al., 1997; Nokelainen et al., 2005; Lim et al., 2012). Now that thinkLets are so well developed, applying it to online collaborative learning can be a promising way to improve learning and collaboration quality. This paper filled the gap between thinkLets and online collaborative learning and may provide new ideas and inspirations in supporting online collaborative learning.

In this study, some keywords related to satisfaction in online collaborative learning are found in the interviews. It might offer something new for building a satisfaction model specially for the combination of thinkLets and online collaborative learning.

**Limitations and Future Work**

There are several limitations of this study. The sample size of the study can be larger to make comparison more reliable. Moreover, exploratory studies like this to some degree trade off some rigor for real-life situations. In comparison groups, it is hard to ensure that differences in satisfaction response are not caused by a difference of quality in the tool implementation (e.g. better usability of the thinkLet version due to increased implementation effort or different tools leading to different user experiences). Controlled experiments can be conducted to make utility of YST and conduct a more scientific study.

To conclude from the study, using thinkLets integrated in GSS is a way to improve yield shifts and satisfaction responses in online collaborative learning. However, there must be other ways to improve participants' satisfaction. Satisfaction of online collaborative learning is important and decisive in learning quality and effectiveness. Hence, other methods besides thinkLets which foster satisfaction can be introduced to improve online collaborative learning quality. Experiments with larger sample size and better controlled experiments should be conducted under a variety of situations to validate the usefulness of thinkLets in online collaborative learning.

# Conclusion

In this paper, students' satisfaction responses with thinkLets and GSS applied in online collaborative learning was measured by YST, indicating whether the attempt of introducing thinkLets and GSS was a success. This study first took YST into online collaborative learning, then tested it in student groups. According to qualitative and quantitative data analysis and implications, three conclusions are made and listed below:

1. Experimental results are consistent with YST, and satisfaction is higher with thinkLets support.

The result showed that the observed relationships between yield shifts and satisfaction responses were consistent with YST propositions, implying that YST is sound when applied to student groups in online collaborative learning. With the help of thinkLets in GSS, participants feel more satisfied. Therefore, in simulating online collaborative learning, adoption of thinkLets is a valid method to achieve higher online collaborative learning quality. This finding might be useful in supporting online collaborative learning and software design, as well as building futher satisfaction model in the field.

2. ThinkLets help to improve quality, enabling more satisfaction with online collaborative learning.

In the interview data, many keywords are mentioned with high frequency. Consensus building, time-saving, and effectiveness are the top three. On the one hand, keywords mentioned indicate the advantages of thinkLets supporting online collaborative learning. On the other hand, they are possible aspects that should be considered in the quality of online collaborative learning.

3. Different methods can be introduced into online collaborative learning as an attempt.

As it is a success to apply thinkLets to online collaborative learning, other attempts can be made in improving quality of online collaborative learning. Daring to try is a positive element for making progress in practice as well as research.

# Acknowledgment

The authors thank the National Natural Science Foundation of China (No. 71571045, No.71101029), Beijing Higher Education and Teaching Reform Project (No.2015-ms080), the Fundamental Research Funds for the Central Universities in UIBE (No.13YQ08) and UIBE Undergraduate Education and Teaching Research Funds for providing funding for part of this research.

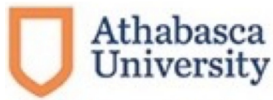

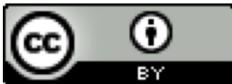